\documentclass[]{aa}
\usepackage{epsfig}
\begin{document}

\title{On the nature of Lithium-rich giant stars\subtitle{Constraints from Beryllium abundances}\thanks{Based on observations collected 
with the VLT/UT2 Kueyen telescope (Paranal Observatory, ESO, Chile) 
using the UVES spectrograph (program ID 69.D-0718A)
}
                            }

 \author{   
            C. H. F. Melo  \inst{1},
            P. de Laverny  \inst{2},
	    N. C. Santos      \inst{3}$^,$\inst{4}
            G. Israelian   \inst{5},
            S. Randich      \inst{6},
            J. D. do Nascimento Jr \inst{7}
      \and  J. R. De Medeiros \inst{7}
  }

   \institute{
      European Southern Observatory, Casilla 19001, Santiago 19, Chile 
    \and
     Observatoire de la C\^ote d'Azur, D\'epartement 
     Cassiop\'ee, UMR 6202, BP 4229, 06304 Nice, France
     \and
     Centro de Astronomia e Astrof\'{\i}ca da Universidade de Lisboa, Observat\'orio
     Astron\'omico de Lisboa, Tapada da Ajuda, 1349-018 Lisboa, Portugal
     \and
     Observatoire de Gen\`eve, 51 ch. des Maillettes, 1290 Sauverny, Switzerland
     \and
     Instituto de Astrof\'{\i}sica de Canarias, 38205 La Laguna, Tenerife, Spain
     \and
     INAF/Osservatorio Astrofisico di Arcetri, Largo E. Fermi 5, 50125 Firenze, Italy
      \and 
    Departamento de F\'{\i}sica,
    Universidade Federal do Rio
    Grande do Norte, 59072-970
    Natal, RN., Brazil
    }

  \offprints{C.H.F. Melo}
  \mail{cmelo@eso.org}

  \date{Received / Accepted}

%\thesaurus{
%            08.01.1;  % Stars: abundances
%            08.05.3;  % Stars: evolution
%            08.09.3;  % Stars: interiors
%            08.12.1;  % Stars: late-type
%}
\abstract{We have derived beryllium abundances for 7 Li-rich giant 
($A(Li) > 1.5$) stars 
and 10 other Li-normal giants, 
with the aim of investigating
the origin of the Lithium in the Li-rich giants. In particular, we
test the predictions of the engulfment scenario proposed by
Siess \& Livio (\cite{siesslivio99}), where the engulfment of a brown dwarf or one or more
giant planets would lead to a simultaneous enrichment of $^7$Li and $^9$Be. 
We show that regardless their nature, none of the stars studied in this paper
were found to have detectable beryllium. Using simple dilution arguments we
show that the engulfment of an external object 
as the {\it sole} source of Li enrichment
is ruled out by
the Li and Be abundance data. The present results favor the idea that Li has been
produced in the interior of the stars by
a Cameron-Fowler process and brought up to the surface by an extra mixing mechanism.
\keywords{ stars:     abundances
           stars:     interiors   --
           stars:     late-type
            }}
  \authorrunning{Melo et al.}
  %\titlerunning{ }
   \titlerunning{On the nature of Lithium-rich giant stars}
\maketitle

\section{Introduction}

The discovery of  lithium excesses in a few low-mass giants in recent years represents one
of the most exciting puzzle for stellar astrophysics. Essentially such stars show
lithium content significantly larger than the values predicted in the framework of
standard stellar evolution, some of them possessing surface lithium content approaching 
the present interstellar medium  value of $A(Li) \sim 3.0$ or
even higher 
(de la Reza \& da Silva~\cite{rezasilva95}, Balachandran et al.~\cite{balaetal00}).
Standard models predict that, after the first dredge-up and dilution due to the
deepening of the convective zone, RGB stars should have a Li abundance 20-60 times
below the initial value (e.g. Iben~\cite{iben67}). However, since low mass stars have
partially destroyed their initial Li on 
the pre-main sequence and on the main sequence, the observed Li abundances
in RGB stars are indeed much lower.

%Such a fact violates the general
%rule that stars approaching the base of the red giant branch (RGB) are expected to have 
%surface lithium abundances from 2 to 4 orders of magnitude below the present
%interstellar medium value, as a result of the deepening of the convective envelope
%(see for instance  Jasniewicz et al. \cite{jasnetal99}; de Laverny et al. \cite{delavetal03}).

In spite of an increasing number of studies with a variety of propositions
(see review by de la Reza \cite{delareza00} and references therein), the root-cause
of these highly abnormal abundances of Li in low-mass red giant stars remains unknown,
adding a new critical question for these undoubtedly very complex physical systems.
Some explanations are related to internal  processes, such as a fresh lithium synthesis
(e.g. Sackmann \& Boothroyd \cite{sacboo99}), or a preservation of the initial 
lithium content (Fekel \& Balachandran \cite{fekbal93}), whereas other
explanations are based on external processes as the
contamination of the stellar external layers by debris of nova ejecta or the 
engulfment of brown dwarfs or planets by the giant star 
(Alexander \cite{alex67}; Brown et al. \cite{brownetal89}; Gratton
\& D'Antona \cite{gradat89}; Siess \& Livio \cite{siesslivio99}).

Particularly interesting have been  the explanations based
on fresh lithium production or planet engulfment, although
recent observational studies have not resulted in any
fully satisfactory conclusion.  Sackmann and Boothroyd
(1999) have shown that, under certain conditions, $^7$Li can
be created in low-mass red giants via the Cameron-Fowler
mechanism, due to extra deep mixing and the associated
cool bottom processing. Following these authors, this fresh
material could account for the excess of lithium observed
in the so-called lithium rich giants.  If deep circulation
is a long-lived, continuous process, these lithium-rich 
stars should be completely devoid of beryllium and boron,
whereas if it occurs in short-lived episodes, beryllium
and boron might be only partially destroyed. 

On the other context, theoretical  predictions by Siess and Livio (\cite{siesslivio99}), assuming that
a planet, brown dwarf or very low mass  star is dissipated at the bottom of the
convective envelope of the giant star, point to several observational  signatures that
accompany the engulfing phenomenon. Among such signatures, there is the ejection of a
shell and a subsequent phase of IR emission, an increase in the light elements
surface content (specially in the $^7$Li, but also $^6$Li, $^9$Be and $^{11}$B), 
potential stellar metallicity enrichment, spin-up of the star because of the deposition
of orbital angular momentum, the possible generation of magnetic fields, and the related
X-ray activity caused by the development of shear at the base of the convective
envelope. 

Concerning the light element enhancement,
Israelian et al. (\cite{israetal01}, \cite{israetal03})
used high-resolution and high signal-to-noise observations to show that
$^6$Li is present in the atmosphere of
the extra-solar planet star HD82943. The authors interpreted the presence of $^6$Li 
as evidence for a planet (or planets) having been engulfed by the parent star.
Observations of Be \textsc{ii} lines in two lithium-rich giants, HD 9746 and HD 112127, 
with the IUE indicated that Be is not probably preserved in these stars 
(De Medeiros et al. \cite{demedetal97}). Such a result seems to indicate that Be 
and consequently the primordial Li were 
depleted in these two Li-rich stars. This suggests that additional Li was produced
in the stellar interior or added from an external source. Castilho et al. (\cite{casetal99}) have
measured Be abundances for two Li-rich giants, concluding that, in such stars, Be is
very depleted in relation to the initial Be abundance value of population I stars.
In the present work we extend the work done in Castilho et al. deriving Be abundances
for all  Li-rich giants visible from the southern hemisphere.
Along with the Li abundances found in the literature, the newly derived
Be abundances allow for a much stronger test of the engulfment scenario
proposed by Siess \& Livio (\cite{siesslivio99}), since the Li enrichment
resulting from an engulfment episode would also be accompanied by
a detectable Be enrichment.

\section{Sample and Observations}

Approximately 1\% of giant stars have anomoulous high Li
abundances compared to the values predicted by the standard models.
Here we have selected 9 Li-rich stars  
which can be observed from the southern hemisphere along with  another 10 Li-normal giants
selected as comparison sample.

%%Table 1 - add the comparison sample, add column indicate the group (Li-rich or comparison)
%%For Li-rich add reference of the discovery paper. 
\begin{table*}
\begin{center}
\caption{Sample stars and Observation log.}
\label{table:sample}
\begin{tabular}{rccccccc}\hline\hline
  HD    &    V  & $(B-V)$ & $\pi$ (mas) & Night & $S/N$ at Be \\ 
\hline
Li-rich giants:\\
 787	&  5.28 & 1.50  &  5.33 &  2002-07-30	   &70  	 \\
 19745  &  9.11 & 1.04  &   --  &  2002-07-30	   &30  	  \\
 30238  &  5.71 & 1.50  &  5.20 &  2002-09-10	   &80  	  \\
 39853  &  5.63 & 1.55  &  4.37 &  2002-09-09	   &30  	  \\
% 65750  &  7.03 & 2.04  &  3.37 &  2002-05-29	   &20  	  \\
 95799  &  7.99 & 1.00  &   --  &  2002-05-21	   &70  	  \\
% 146850 &  6.01 & 1.50  &  3.77 &  2002-06-02	   &70  	  \\
 176588 &  6.89 & 1.72  &  4.25 &  2002-06-01	   &80  	  \\
 183492 &  5.57 & 1.04  & 11.38 &  2002-06-02	   &80  	  \\
 217352 &  7.16 & 1.15  &  5.11 &  2002-07-31	   &50  	  \\
 219025 &  7.68 & 1.21  &  3.25 &  2002-07-29	   &30  	  \\
\hline
Li-normal stars:\\
%787	&  5.28 & 1.50  &  5.33 & \\
360	& 5.99	&1.03	& 9.79	&  2002-08-04	   &90  	 \\
1522	&3.56	&1.21	&11.26	&  2002-08-04	   &60  	 \\
4128	&2.04	&1.02	&34.04	&  2002-08-04	   &220 	 \\
5437	&5.35	&1.51	&6.24	&  2002-09-10	   &80  	 \\
61772	&4.98	&1.54	&4.82	&  2002-04-18	   &60  	 \\
61935	&3.94	&1.02	&22.61	&  2002-04-18	   &130 	 \\
95272	&4.08	&1.08	&18.71	&  2002-04-18	   &100 	 \\
105707	&3.02	&1.33	&10.75	&  2002-04-29	   &70  	 \\
126271	&6.19	&1.21	&8.97	&  2002-05-31	   &100 	 \\
220321	&3.96	&1.08	&20.14	&  2002-06-13	   &80  	 \\

\hline
\end{tabular}
\end{center}
\end{table*}

Observations of the Be \textsc{ii} doublet at 3131\AA\ for the two groups were carried out in
service mode using the UVES spectrograph (Dekker et al. \cite{dekkeretal00}) attached to
VLT/UT2 Kueyen telescope at ESO, Chile. 
UVES was operated in the blue-arm which is equipped with a EEV 2048x4102 CCD.
Using the cross-disperser \#1, the blue-arm covers the spectral range from 
$\sim3020$ to $\sim3880$ \AA\.
The spectral region around the Be \textsc{II} doublet is
heavily crowded. Therefore, a spectral resolution $R$ better than about  30000-40000 is
mandatory (Garc\'{\i}a L\'opez et al. \cite{garetal95}). In order to have a high
signal-to-noise (around 100) and a high-resolution spectrum even during poor seeing
conditions, the image slicer was inserted (Dekker et al. \cite{dekkeretal02}). The
obtained spectra have a resolution of around 70000 and a $S/N$ ratios about
 70-80 in the region of the Be \textsc{ii} doublet for
all but three stars which have a $S/N\sim20-30$. Apparent visual magnitudes,
$(B-V)$, parallaxes and $A(Li)$ taken from the literature 
along with the $S/N$ ratios obtained  are listed in Table \ref{table:sample} for both samples. 
The data
collected was reduced using IRAF\footnote{IRAF is distributed by National Optical Astronomy 
Observatories, operated by the Association of Universities for 
Research in Astronomy, Inc., under contract with the National 
Science Foundation, U.S.A.} (echelle package) to carry out the standard  echelle
spectrum reduction steps, namely, bias and background subtraction, flat-field
correction, spectrum extraction and wavelength calibration. The wavelength calibration
was done using as reference a ThAr lamp spectrum taken in the morning of the following
day.

\section{Be abundance calculations}
\subsection{Stellar parameters}
For the program stars, effective temperatures have been estimated
from $(B-V)$ photometry found in the Hipparcos catalog (Perryman et al.
\cite{perrymanetal97}) and the calibrations of Alonso et al. (\cite{alonsoetal99})
and Houdashelt et al. (\cite{houdasheltetal00}).
We also considered $(b-y)$ colors, when available,
from Hauck \& Mermilliod (\cite{hauckmermio98})
and the calibration of Alonso et al. (\cite{alonsoetal99}). In the 
color-$T_{\rm eff}$ calibrations,
we assumed a solar-metallicity for all the stars.
For comparison, temperatures were derived with the different
methods/colors being in rather good agreement, implying that
they should not be affected by too large systematic errors
and that [Fe/H] should not be significantly different from solar. 
Adopted effective temperatures are the ones computed using the  $(B-V)-T_{eff}$ 
calibration of Alonso et al. (\cite{alonsoetal99}).

\begin{figure}[t]
\centering
\resizebox{\hsize}{!}{\includegraphics[width=10cm,angle=-90]{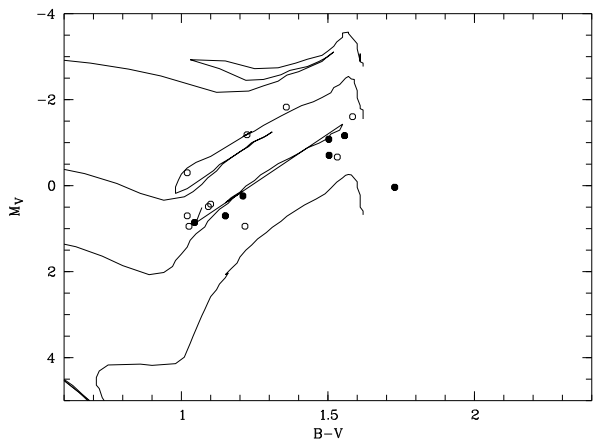}}
\caption{Color-magnitude diagram of the observed sample. Li-normal and Li-rich giants are seen as heavy
and light gray points, respectively. Overplotted are isochrones 
from Bertelli et al. (\cite{bertellietal94}) for (from the bottom to top) log age=10, 9, 8.6 and 8 yrs
corresponding to masses on the RGB of 1, 1.9-1.95 and 4-5.3 solar masses.}
\label{fig:cmd}
\end{figure}

Parallax-based stellar gravities have been estimated using 
parallaxes and visual magnitudes taken from the Hipparcos catalog  with the
bolometric corrections being computed 
using the Flower~(\cite{flower96}) $T_{eff}-BC$ calibration and effective temperatures
computed as described above. Also, a $M_{v,\odot}=4.83$
and a $BC_\odot=-0.12$ were adopted. Stellar masses were estimated 
by comparing the position of our sample stars in the color--magnitude diagram
with the evolutionary tracks from Bertelli et al. (\cite{bertellietal94}) computed using
solar metallicity as shown in Fig.~\ref{fig:cmd}. Surface gravities were computed using Eq. 3 of
Nissen et al. (\cite{nissenetal97}).

%In the revised version of the CM diagram, stars in black
% are Li-normal stars, while stars in red are the Li-rich giants.
% Isochrones correspond (from the bottom) to log age=10, 9, 8.6 and 8 yrs.
% Corresponding masses of stars on the RGB are: 1, 1.9-1.95, 4-5.3 M_sun.

%Most of the stars have masses between 1 and 2 $M_\odot$, 
%a few of them having $M \sim 2M_\odot$ and another small fraction are
%more massive. Thus the $\log g$ was computed assuming a mean stellar mass of 1.5$M_\odot$ for
%all stars in the sample. Since $\log g \propto \log M$, 
%the impact of using a mean value for the stellar mass is small. For instance, if masses are wrong
%but a factor of 2, the $\log g$ will wrong by 0.3 dex. In our case, most of the masses are
%concentrated between 1.5-2.0$M_\odot$, thus in the worst case the $\log g$ used here is
%wrong by 0.12 dex. 

%In the case a par. is needed to explain the error on logg
%
%Errors in parallax-based stellar gravities are dominated mainly by the uncertaints in the 
%mass determination based on the evolutionary tracks, 
%$BC$ and $T_{eff}$ determination along with the error on the measurement
%of the parallax itself. Nissen et al. (\cite{nissenetal97}) 
%estimate that the total error on the parallaxe-based gravities is of 0.3 dex.

In the spectral analysis, we used these parallax-based gravities
when the error on the parallax was smaller than 15\%.
For the remaining stars without a parallax or those whose mass could
not be well determined (HD176588), we assumed a gravity
typical for giant stars $\log g$ = 2.0. 

As for the metallicity, we considered mean values from
the literature. We also assumed for
all the stars a depth independent microturbulence velocity of 2.0~km s$^{-1}$
with an uncertainty of about 1.0 km s$^{-1}$. {A macro-turbulent velocity value of 2.0~km s$^{-1}$ was also
taken for all stars. Changing this latter by $\pm$2~km s$^{-1}$ would not change
any of the results presented below.}
Rotational velocities from De Medeiros \& Mayor (\cite{medmay99}) were
considered and $V_{\rm sin i} = 1$~km s$^{-1}$ was used for the 
slower rotators.

The final values for the stellar masses, $T_{\rm eff}$ and $\log g$ are given in
Table~\ref{tab:param}. We estimate that the accuracy of our derived 
$T_{\rm eff}$ and $\log g$ are
respectively $\sim100$ K and 0.3 dex and about 1km.s$^{-1}$ 
for the microturbulence.

\begin{table*}
\caption[]{Atmospheric parameters and derived abundances for the program 
stars. Asterisks indicate guess values for the gravity or the metallicity.
For stars with very large rotational velocities, we were unable to derive
abundances of O, Mn and Tm.}
\label{tab:param}
\begin{flushleft}
\begin{tabular}{rcccccccccc}
\hline\hline
$HD$  & ${T_{\rm eff}}(K)$ & log\, $g$ & $M/M_\odot$ &
$\rm [Fe/H]$ & $V_{\rm sin i}({\rm
km\,s^{-1}})$ & $\rm [O/H]$ & $\rm [Mn/H]$ & $\rm [Tm/H]$ & $\rm A(Be)^{\dagger}$ & $A(Li)$  \\
\hline
Li-rich giants:\\
   787  & 3990 &  1.3 & 1.9  & +0.00 & 1.9 & +0.0 & -0.4 & -1.0 & no detection  & 1.8$^a$\\
 19745  & 4730 &2.0(*)&      & +0.10 & 1.0 & +0.6 & +0.3 & +0.5 & no detection  & 4.1$^b$\\
 30238  & 3990 &  1.4 & 1.9  & +0.00 & 1.0 & -0.2 & +0.0 & +0.0 & no detection  & 0.8$^c$         \\
 39853  & 3920 &  1.1 & 1.6  & -0.40 & 1.0 & +0.0 & +0.0 & +0.0 & no detection  & 2.8$^d$\\
% 65750  & \multicolumn{10}{l}{Not observed}		\\
 95799  & 4800 &2.0(*)&      & -0.11 & 1.0 & +0.8 & +0.8 & +0.8 & no detection  & 3.2$^e$\\
%146850  & \multicolumn{10}{l}{Bad fit}		\\
176588  & 3800 &2.0(*)&      & +0.00 & 1.0 & -0.2 & -0.3 & -0.4 & no detection  & 1.1$^c$           \\
183492  & 4720 &  2.6 & 2.0  & +0.00 & 1.0 & +0.6 & +0.6 & +1.1 & no detection  & 2.0$^a$\\
217352  & \multicolumn{10}{l}{too large $V\sin i$}		\\
219025  & \multicolumn{10}{l}{too large $V\sin i$}		\\
\hline
Li-normal stars:\\
   360  & 4750 &  2.7 & 2.1   & -0.20 & 1.0 & +0.5 & +0.2 & +1.0 & no detection  & 0.2$^a$\\
  1522  & 4400 &  1.8 & 3.2   & +0.20 & 1.0 & -0.3 & +0.0 & +0.0 & no detection  & $<0.0$$^a$\\
  4128  & 4760 &  2.4 & 3.2   & +0.15 & 1.0 & -0.2 & +0.0 & +0.4 & no detection  & $<0.2$$^a$\\
  5437  & 3950 &  1.3 & 1.4   & -0.20 & 1.0 & +0.2 & -0.2 & +0.0 & no detection  & $0.1$$^a$\\
 61772  & 3880 &  1.0 & 1.9   & +0.10 & 1.0 & -0.2 & -0.4 & -0.4 & no detection  & -0.5:$^a$\\
 61935  & 4760 &  2.7 & 2.4   & +0.00 & 1.0 & +0.2 & +0.3 & +1.0 & no detection  & 0.2$^a$\\
 95272  & 4620 &  2.4 & 2.1   & +0.00 & 1.0 & +0.4 & +0.2 & +0.9 & no detection  & $<0.0$$^a$\\
105707  & 4190 &  1.4 & 3.2   & +0.10 & 1.0 & +0.2 & +0.3 & +0.0 & no detection  & 0.8$^a$\\
126271  & 4410 &  2.3 & 1.4   & +0.00 & 1.0 & +0.1 & +0.1 & +0.6 & no detection  & $<-0.4$$^a$\\
220321  & 4630 &  2.4 & 2.1   & -0.30 & 1.0 & +0.6 & +0.6 & +1.0 & no detection  & $<-0.2$$^a$\\
\hline
\end{tabular}
\end{flushleft}
$^a$Brown et al. (\cite{brownetal89}) $^b$de la Reza \& da Silva (\cite{rezasilva95})
$^c$Castilho et al. (\cite{casetal00})$^d$Gratton \& D'Antona (\cite{gradat89})
$^e$Luck (\cite{luck94})\\
\newline
$^\dagger$ No detection is compatible with $A(Be)<-5$.
\end{table*}

\subsection{LTE chemical analysis of the program stars}

The chemical abundances were derived by fitting synthetic spectra to the
observed profiles with the set of stellar parameters given in
Table~\ref{tab:param}.
LTE synthetic spectra were computed using a new version of the code MOOG
written by C. Sneden (\cite{sneden73})\footnote{
http://verdi.as.utexas.edu/moog.html} and
model atmospheres were interpolated
in a grid of Kurucz (\cite{kurucz93}) ATLAS9. 

\begin{figure*}[t]
\centering
\resizebox{\hsize}{!}{\includegraphics[width=10cm,angle=-90]{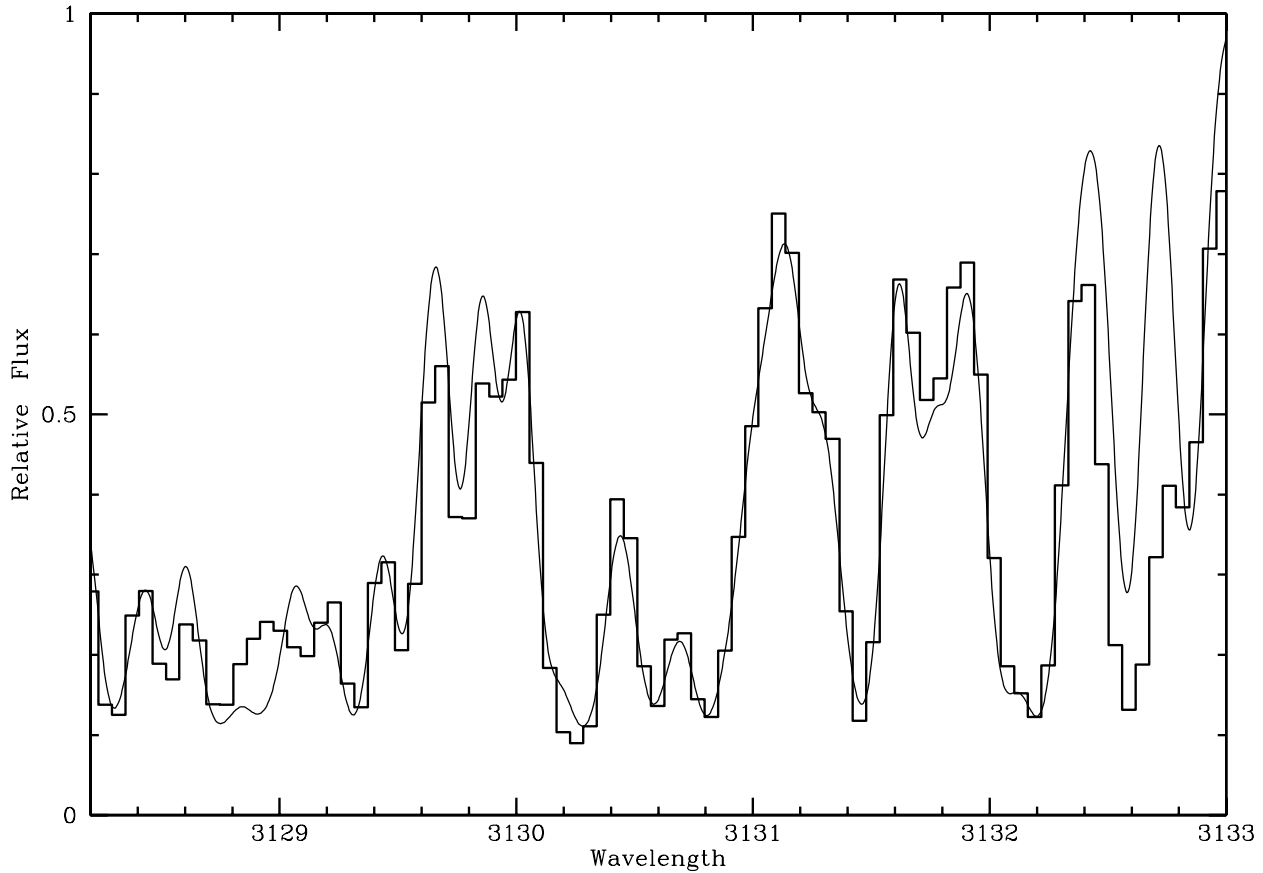}}
\caption{Spectrum of Arcturus and its synthesis used in the calibration of the 
linelist provided by Garcia Lopez et al. (\cite{garetal95}) specially
for cooler-stars. The oscillator strength of some lines between 3130 and 3132~\AA \, were corrected in order to
better fit the Arcturus spectrum assuming the abundances reported by Peterson et
al. (1993)}
\label{fig:arcturus}
\end{figure*}

We used the linelist provided by Garcia Lopez et al. (\cite{garetal95}) specially
built for solar-type stars. In order to test it for cooler stars
as those studied in this work, we calibrated this linelist with the spectrum of Arcturus, 
a K1 giant with no detectable Li or Be in its surface. We further
assumed the stellar parameters and abundances
derived by Peterson et al. (\cite{petersonetal93}).
The oscillator strength of some lines between 3130 and 3132~\AA \, were corrected in order to
better fit the Arcturus spectrum assuming the abundances reported by Peterson et
al. (1993). All the gf-values of the OH lines were decreased by 0.4~dex and
we adopted $gf=6\times 10^{-2}$ for CH 3130.648,
$gf=0.1$ for Cr \textsc{i} 3131.212 and  $gf=8.61\times 10^{-3}$ for
Co \textsc{i} 3131.825.
For the calibration of the Zr \textsc{I} 3131.109 and Mn \textsc{I} 3131.037
lines, we first derived the abundances of Zr and Mn from the Arcturus atlas
using line data from Feltzing \& Gustafsson (\cite{feltzinggus98}). We found [Zr/Fe]=-1.0
and [Mn/Fe]=-0.17 and derived $gf$-values equal to 3.98$\times 10^{-2}$
and 0.296, respectively. 
 For some giant stars, it was also found that the Tm \textsc{ii} 3131.255 may play an important
role. But we were unable to derive the abundance of Tm in Arcturus.
We therefore scaled the abundance of Tm with Fe (as done by Peterson et al., \cite{petersonetal93})
and checked that the gf-value of this line found in the original linelist
leads to a good fit.
Finally, it was found unnecessary to add any {\it ad hoc} Fe \textsc{i} lines
as proposed by Castilho et al. (\cite{casetal99}) at 3130.995~\AA \, and 
Primas et al. (\cite{primasetal97}) at 3131.043~\AA.
The final spectral synthesis of Arcturus is shown in Fig.~\ref{fig:arcturus}.

From the estimated stellar parameters, we first derived abundances
of O, Mn and Tm using the OH lines, the Mn \textsc{i} line at 3131.037\AA\ and 
Tm \textsc{ii} line 3113.255\AA. In a second iteration, the
 Be was then determined. They are reported in
Table~\ref{tab:param}. When comparing observations with our synthetic spectra, 
it was sometimes found necessary to add a small amount of veiling (about 10\%)
to the synthetic spectra in order to improve the fits of the cores of the
most saturated lines. We attribute this
either to an incorrect background subtraction\footnote{Since we
used the image slicer, the
flux-per-pixel is quite low in our spectra,
and this may complicate the background light subtraction
procedure.} or to less reliable models of atmospheres
for cool giants leading to an imperfect treatment of the
line cores. In Santos et al. (\cite{santosetal02}) this problem was
not found for cool dwarfs, although we have not
studied in this latter work any star with $T_{eff}$ lower
than 5200K. In any case, and since the Be abundances were
estimated from the redder BeII line (not affected by
saturated lines), a few tests have shown that this
does not change any of the results presented here
regarding the Be abundances.

{Alternatively, this veiling could be related to the solar missing opacity problem (Balachandran \& Bell, \cite{balabell98}).
However, it is not clear if one can extrapolate this issue for the domain of cool giant stars. Also, it may be that this
missing opacity is not needed even for the Sun (e.g., Allende Prieto \& Lambert \cite{allelam00}).}

{
We have tested the sensitivity of our derived Be abundances to changes in surface gravity ($\pm0.5$ dex),
effective temperature ($\pm150$ K) and macro-turbulence ($\pm 2 $km s$^{-1}$). The results show that for
such a low Be levels, no changes are found. In all cases, the final abundances are always compatible with no detection
of Be.} Other possible sources of errors in the derived Be abundances are well
discussed in Garc\'{\i}a L\'opez (\cite{garetal95}) and 
Santos et al. (\cite{santosetal04b}). We refer to these works for further
details.

%I woud like to say that:
%(Results) Present the abundace results: zero Be
%(Discussion) Indicar a ordem de grandeza do efeito que estamos procurando (em termos de A(Be)). 
%
%	- De acordo com o modelo de diluicao quanto Li e Be se esperaria
%	- Siess & Livio: quanto Be viria de um suposto engulfment?
%
%Dizer que a resolucao e sensibilidade sao suficientes para detectar abundancias de tanto que
%que correspondem a um engulfment de uma estrela ou N planetas de tal massa e tal composicao
%
%Mencionar que a origem do execesso de Li continua intrigante mas que o fato de nao haver Be favorece
%fonte interna de Li. Mas que o excesso de IR pode sugerir engulfment. 6Li pode ajudar a concluir?
%
%\subsection{Comparison with previous work}

\section{Results and discussion}

\subsection{Be is fully depleted}

\begin{figure}[t]
%\begin{tabular}{c}
\centering
%Li rich 19745 et 95799 
%li-nroaml 360/95272/126271 
\resizebox{\hsize}{!}{\includegraphics[width=10cm,angle=-90]{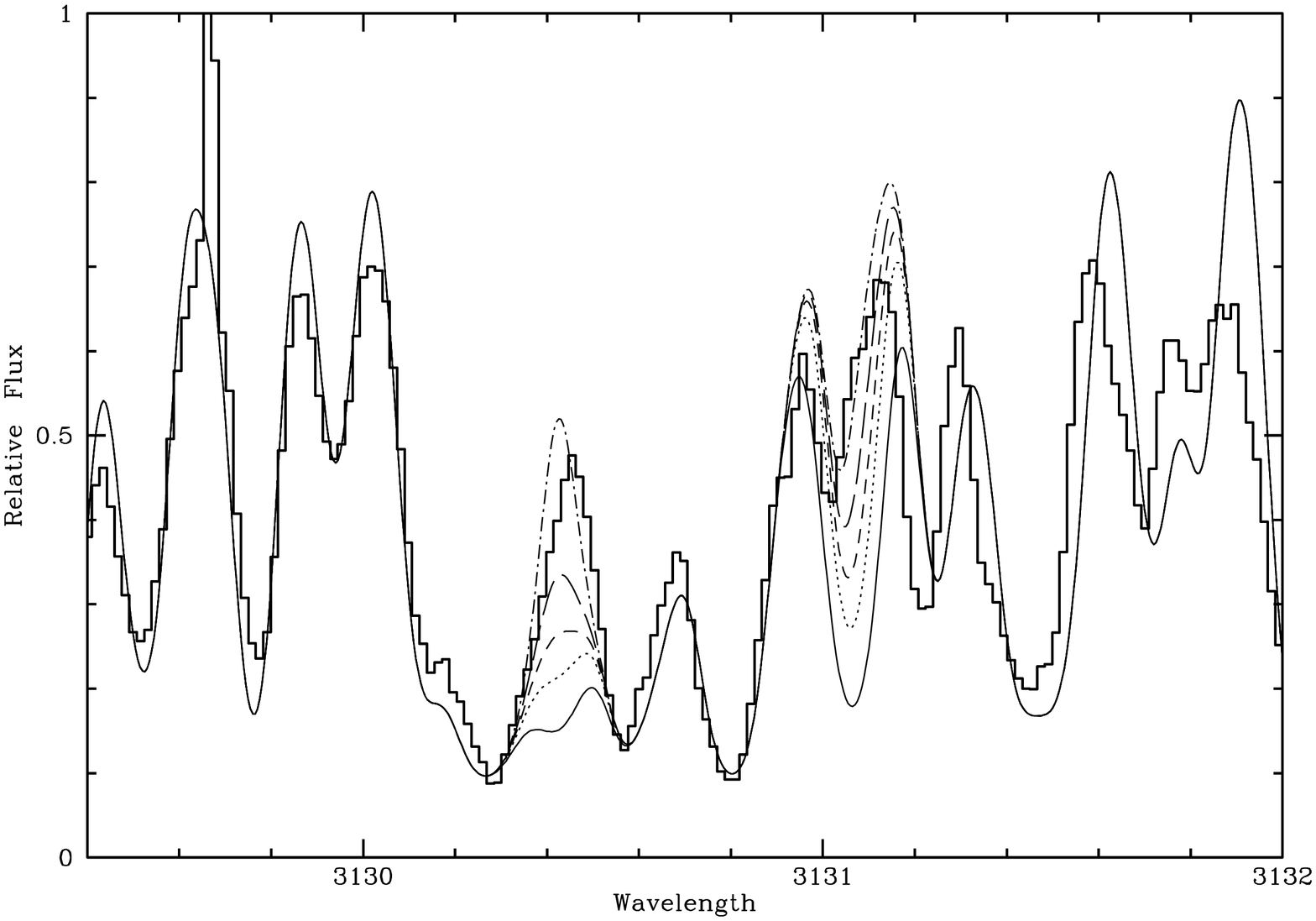}}\\[-0.5cm]
\resizebox{\hsize}{!}{\includegraphics[width=10cm,angle=-90]{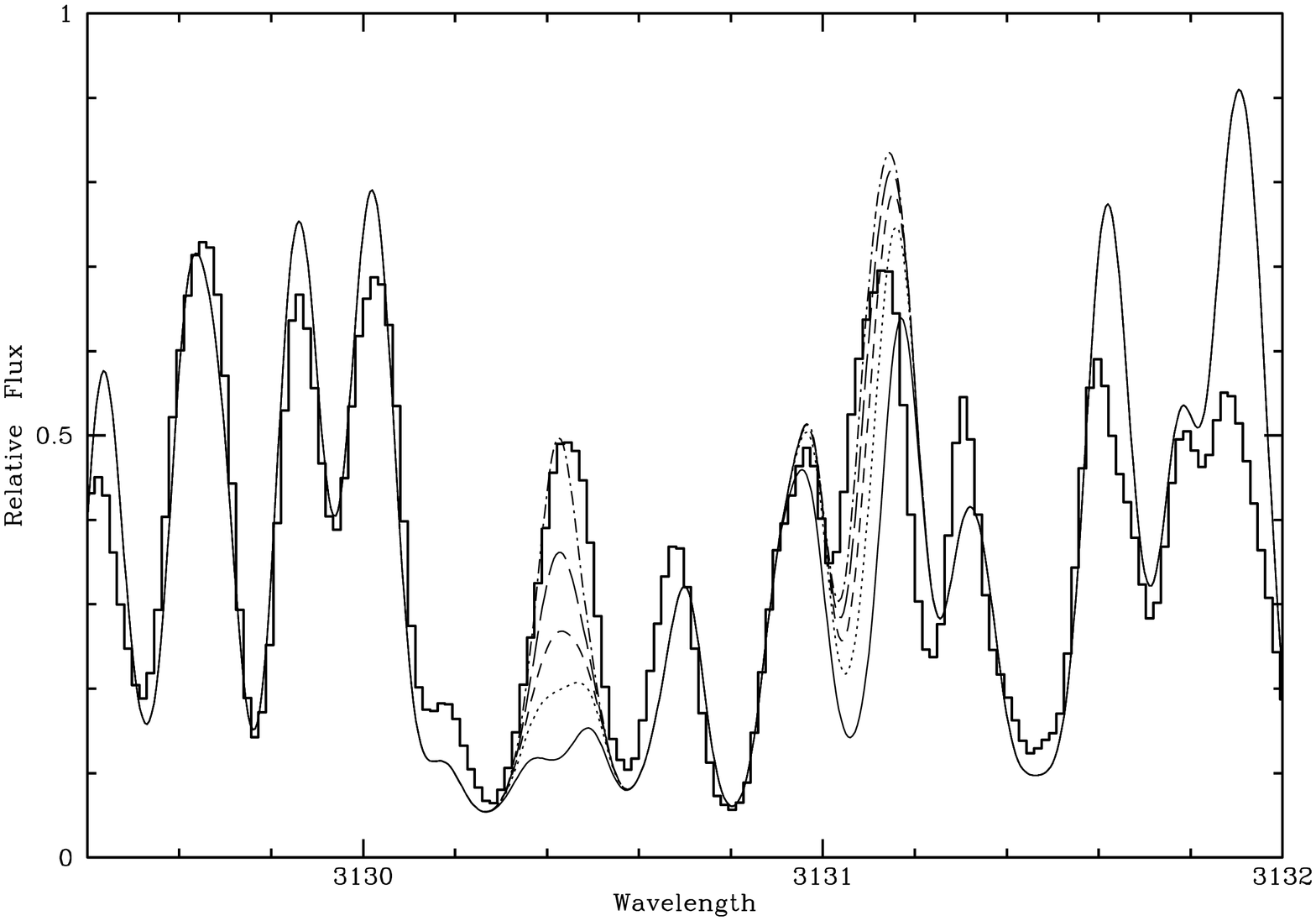}}\\
%\resizebox{0.47\hsize}{!}{\includegraphics[width=10cm,angle=-90]{plots/fig360_FigPubli.ps}}&
%\resizebox{0.47\hsize}{!}{\includegraphics[width=10cm,angle=-90]{plots/fig126271_FigPubli.ps}}\\
%\end{tabular}
\caption{Example of spectral syntheses (continuous lines) and observed spectra (histogram)
in the Be II line region for 2 Li-rich stars HD19745 and HD95799.
In all panels, the upper to lower syntheses corresponds to $A(Be)$ of 
0.0, -1.0, -1.5, -2.0 and -5.0}
\label{fig:synth}
\end{figure}

The main results from the present observations are reported in Table~\ref{tab:param},
where in addition to the beryllium abundances $A(Be)$, [O/H], [Mn/H] and
[Tm/H] are also presented for the Li-rich giants and the Li-normal giants. 
%In the column status, Li-rich indicates the so called
%lithium-rich giants and normal Li those giant stars with lithium abundance following the
%general rule of dilution for their spectral types. 

A first important and striking result
concerns the very low values of
$A(Be)$.
%\footnote{$A(Be)=[Be/Fe]+A(Be)_\odot+[Fe/H]$
%
%where $A(Be)_\odot=1.15$ (Chmielewski et al. \cite{andersgrevesse89})}. 
Beryllium is essentially
absent in the atmosphere of
all stars presented in Table~\ref{tab:param}, regardless their normal or
lithium-rich behavior. In Fig.~\ref{fig:synth} we show the spectral synthesis
for two Li-rich and for two Li-normal giants.
Given the quality of the spectral syntheses seen in Fig.~\ref{fig:synth} 
we can promptly see that our linelist fits well  
the entire region around the Be \textsc{ii} which excludes a poor spectral synthesis as an explanation for
the low values of $A(Be)$.

%Falar que in spite of the value that one can derive with different set of atm. param. 
%THERE'S NO Be THERE!
%Falar aqui do paper de Bruno?

The fact that no Be was found in the Li-rich stars is intriguing. 
According to classical stellar evolution models, 
Li and Be depletions  are not expected to take place during the main-sequence, since the
survival zones for both elements are deeper than the depth of the convective zones of F-G
dwarfs (e.g., review by Deliyannis et al. \cite{deletal00}). During the post-main-sequence
evolution, as the convective zone deepens, the stellar material "rich" in light elements (Li, Be and B) is mixed
with material with no such elements, so the surface content is decreased by dilution.
Boesgaard et al. (\cite{boesetal77}) compared both Li and Be abundances of four giants in the Hyades to
the abundance values measured among the dwarfs in same cluster. They found that
the giants are deficient in Li by a factor of 110 whereas Be is depleted by a factor between 20-70
as compared to the dwarfs in cluster.

Although the standard models can provide some insight into the light element depletion inside the stars, 
it is a known fact that they are not able to explain
most of the observed behaviors of Li and Be (e.g. Stephens et al.~\cite{steetal97}).
This fact leads to the exploration of alternative depletion mechanisms acting already on the
main-sequence.
%In spite of the lack of observational constraint, 
%different mechanisms have been recently proposed to explain the observed 
%Li and Be depletion pattern.
For instance, simultaneous Li and Be
depletion is foreseen in models with diffusion for stars cooler than about 6600 K whereas
models based on mass loss predict  
a flat Be versus
Li distribution. In models with slow mixing, Li depletion should be
more efficient than Be, but a correlation between Be and Li is predicted (see review by Delyannis
et al. \cite{deletal00} and references therein). 

%In this context, giant stars presenting a 
%clear signature of Li depletion should
%present also signatures of depletion in Be. Such a fact is observed in the Pleiades
%giants, where both Be and Li  are deficient in giants in comparison with their abundances in
%the dwarf stars (Boesgaard et al. 1977). 
%
%On the other hand, the so called lithium-rich
%giants should present also an excess in Be if this element was preserved during evolution
%off the main sequence. If, on the contrary, lithium excess results from fresh Li synthesized
%in the stellar interior, one should not expect for enhanced Be abundances in the stellar
%atmospheres; from external contamination, for example the engulfment of planets, brown
%dwarfs of very low mass stars, an excess of Be should be also expected.

%Point: Be *might* have depleted in the MS depending on the Teff. In worst case,
%These stars experienced a strong Be depletion/dilution and such must have been the Li one.

Recent observational
work devoted to the measurement of Li and Be abundances for main-sequence dwarfs have
shown that {\it both} elements are indeed depleted during the main-sequence 
(Stephens et al. \cite{steetal97};
Santos et al. \cite{santosetal02}, \cite{santosetal04}). 
In particular, Santos et al. (\cite{santosetal02}, \cite{santosetal04})  have shown that Be
abundances (and depletion rates) seem to be a function of the effective temperature with a maximum close to the
meteoritic value ($A(Be)=1.42$, Anders \& Grevesse~\cite{andersgrevesse89}) near
$T_{eff}\sim6100$K, decreasing both towards higher temperatures (Be-gap for F stars), and
lower temperatures. A similar trend is also observed for the Li abundances.
Although Li and Be depletion rates does not necessarily happen at the same rate
(Santos et al. \cite{santosetal02}, \cite{santosetal04}, although see Boesgaar 
et al. \cite{boesetal01} for a counter example of correlated Li and Be depletion), 
the fact that i) both elements are depleted already on the main-sequence and 
that, as a general rule, ii) Li depletion takes place {\it much faster} than Be seem to be
undeniable. 

For stars on the blue side of the Li gap (i.e., $T_{eff}\ga 7000$ K or $M\ga M_\odot$) no much information 
on the light element abundances is available. In
a series of papers Burkhart \& Coupry (\cite{burcou97}, \cite{burcou98}, \cite{burcou00})
measured Li abundances of normal A- and Am-stars (slowly rotating magnetic A stars) in different open clusters. 
They show that, for each studied cluster,
the Li abundances of the normal A stars set the maximum  value for the cluster. Normal A
stars are not believed to deplete Li either during the pre-main sequence or during the main-sequence phase,
in addition, being fast rotators they are supposed to be less affected by 
any depletion caused by settling or separation process. Therefore, giant stars in this mass regime 
had probably not experienced Li depletion during their PMS and MS phase.
%Therefore,
%in the light of the Li and Be depletion patterns observed in the main-sequence,
%the total absence of Be in the Li-rich giants cannot be seen as 
%a direct outcome the main-sequence depletion history
%since a combined depletion of both elements
%(not necessarily correlated) which is expected to occur.
In conclusion, the complete
absence of Be in the atmospheres of Li-rich giants in our sample cannot
be ascribed to their MS depletion histories, due to the 
facts that i) a combined depletion
of both elements is expected to occur
and that ii) Li is destroyed faster than Be. For more massive stars at the blue side of the Li gap, no 
substantial depletion is expected.

%{\bf (We might consider add something about the fact that most of the MS Li/Be works is about
%F-M stars. Some of the Li-rich giants have $M>2M_\odot$ which
%corresponds to A-type stars, i.e., hotter than the Li and Be gap in the main sequence. 
%Shall we simply say that no depletion is expected in this mass regime?}

%Ou seja, tanto to ponto de vista teorico como observacional nao se espera encontrar
%Li sem Be!

%\begin{figure}[t]
%\vspace{.2in}
%\centerline{\psfig{figure=Be_Giants_fb0.ps,width=3.8truein,height=3.8truein}
%\hskip 0.1in}
%\caption[]{ The IRAS source position analysis.
%}
%\label{Fig0}
%\end{figure}

%\begin{figure}[t]
%\vspace{.2in}
%\centerline{\psfig{figure=Be_Giants_fb1C.ps,width=3.8truein,height=3.8truein}
%\hskip 0.1in}
%\caption[]{ The IRAS $[12]-[25]$ vs. $[25]-[60]$
%color--color diagram
%for Li rich (which are represented by solid circles) and
%normal Li stars. The rectangular TTauri box denotes
%the color
%limits used by Harris et al. (1988) to define T Tau-like
%sources and he rectangular Giant box denotes the color
%limits for most of the evolved stars (Ivezi\'c \&  Elitzur
%2000).
%}
%\label{Fig1}
%\end{figure}
\begin{figure}[t]
%\begin{tabular}{c}
\centering
%Li rich 19745 et 95799 
%li-nroaml 360/95272/126271 
%\resizebox{\hsize}{!}{\includegraphics[width=10cm,angle=-90]{plots/fig19745_FigPubli.ps}}\\
%\resizebox{\hsize}{!}{\includegraphics[width=10cm,angle=-90]{plots/fig95799_FigPubli.ps}}\\
\resizebox{\hsize}{!}{\includegraphics[width=10cm,angle=-90]{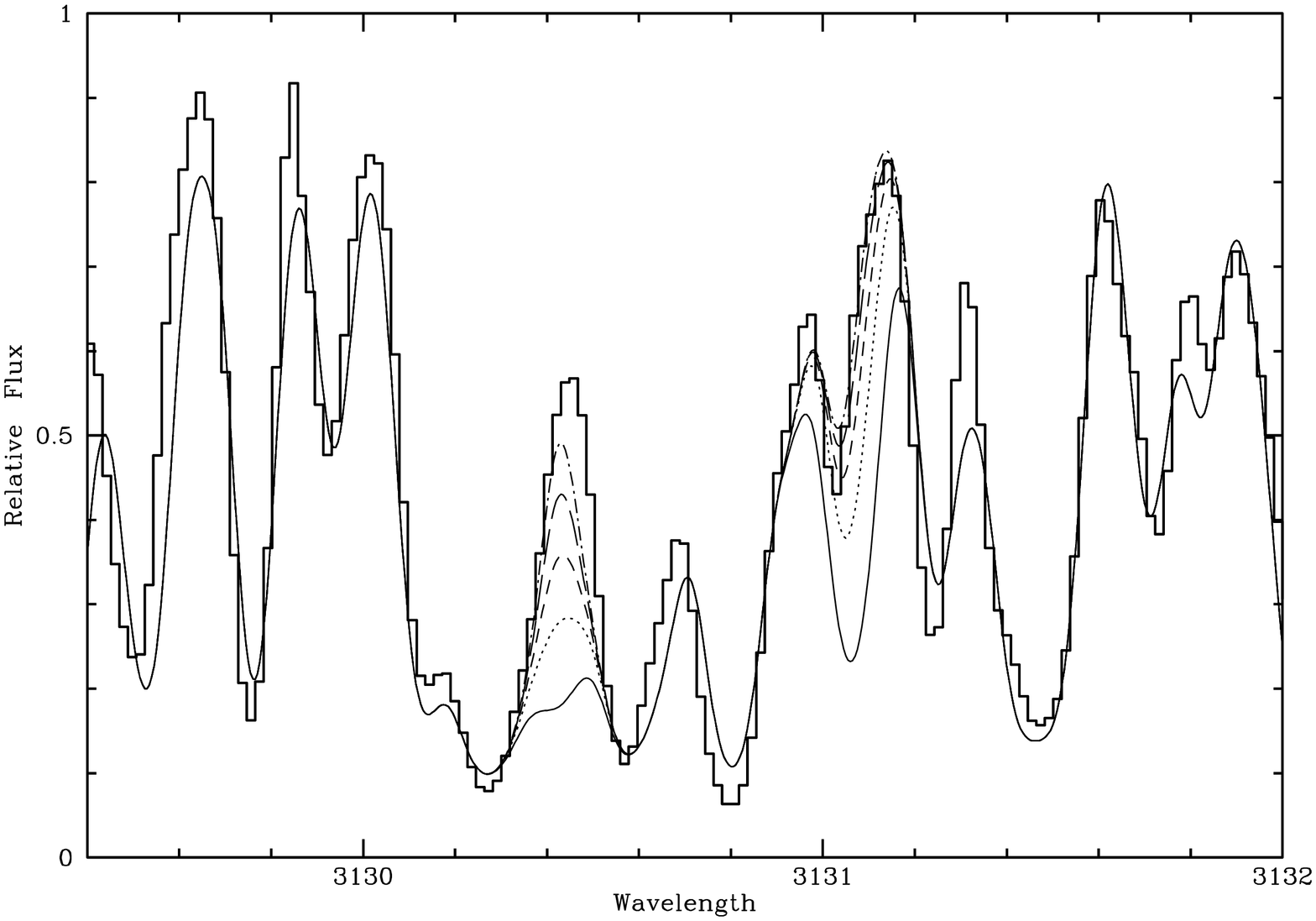}}\\[-0.5cm]
\resizebox{\hsize}{!}{\includegraphics[width=10cm,angle=-90]{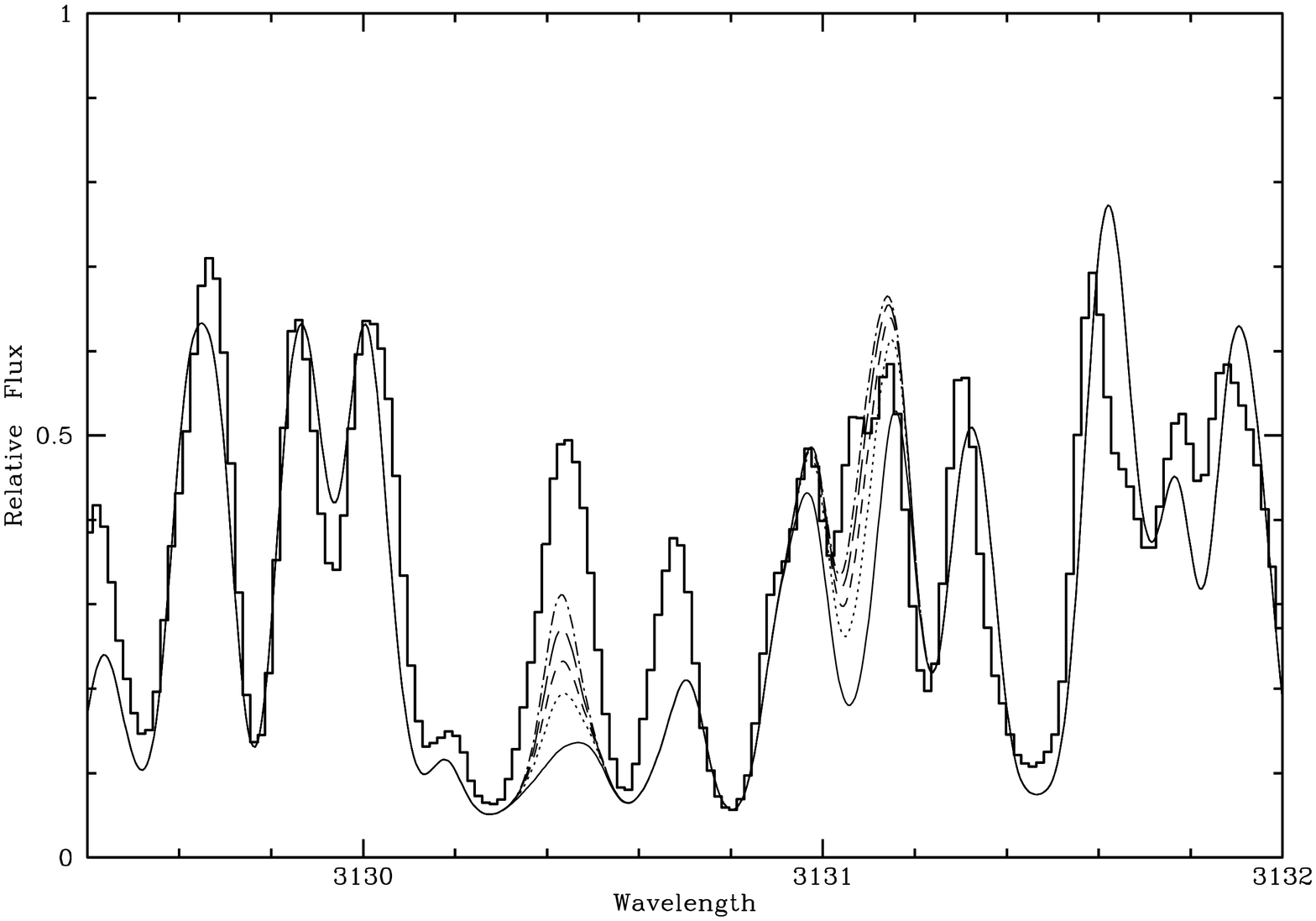}}\\
%\end{tabular}
\caption{Same as Fig.\ref{fig:synth} for the 2 Li-normal stars
HD360 and HD126217.}
\label{fig:synth2}
\end{figure}

\subsection{The origin of the Li in the Li-rich giants}

The possible origin of the Li in the Li-rich giants has been largely discussed in the
literature for many years now (see e.g. Charbonnel \& Balachandran \cite{chabal00},
de la Reza \cite{delareza00} and reference therein). 
So far, none of the proposed scenarios have been found to explain the the existence of the Li-rich giants.
Nonetheless, the internal mixing and convective scenarios seem appealing giving 
our results for the Be abundances.
These scenarios are based in the Cameron \& Fowler (\cite{cameronfowler71}) process in
which $^7$Be is produced in the hot internal layers corresponding to the H-burning
zone through the reaction $^3$He$(\alpha,\gamma)^7$Be. 
For giants with masses between 3-6$M_\odot$, this region (i.e., the H-burning zone) is close to the
the internal base of convective zone, therefore $^7$Be produced by the reaction above  
is promptly transported to the outer layers where it is finally transformed into $^7$Li by the 
reaction $^7$Be$(e^-,\nu)^7$Li. This mechanism is known as the Hot Bottom Burning
proposed by Sackmann \& Boothroyd (\cite{sacboo92}).
For less massive giants $<2.5M_\odot$ the internal base of the convective zone does not reach the H-burning zone,
thus the $^3$He from the envelope cannot be converted into Be. In order to solve this problem,
Sackmann \& Boothroyd (\cite{sacboo99}) proposed the Cool Bottom Processing model, in which 
an {\it ad-hoc} two-stream
conveyor-belt is responsible for circulating the $^3$He-rich envelope of the red giant to the
H-burning zone producing fresh $^7$Be which is transported to the external layers and
transformed into $^7$Li again by the reaction $^7$Be$(e^-,\nu)^7$Li.

%Por um lado nossos resultados parcem colar bem com O CBP: Desenvolver
On the one hand, the CBP model described above seem to qualitatively explain our results. According
to Sackmann \& Boothroyd (\cite{sacboo99}), if the deep mixing is continuous and long-lived then
the star must show a Li enrichment due to fresh Li brought to the surface and be
{\it completely} devoid of beryllium and boron.
Charbonnel \& Balachandran (\cite{chabal00}) on the other hand argue that, if such a long-lived mixing
episode occurs this must be reflected in the values of $^{12}$C/$^{13}$C which should
continuously decrease along the RGB reaching values much lower than the standard values resulting from 
the first dredge-up, which is not observed.

\begin{figure*}[t]
\begin{tabular}{cc}
\centering
\resizebox{0.47\hsize}{!}{\includegraphics[width=10cm,angle=-90]{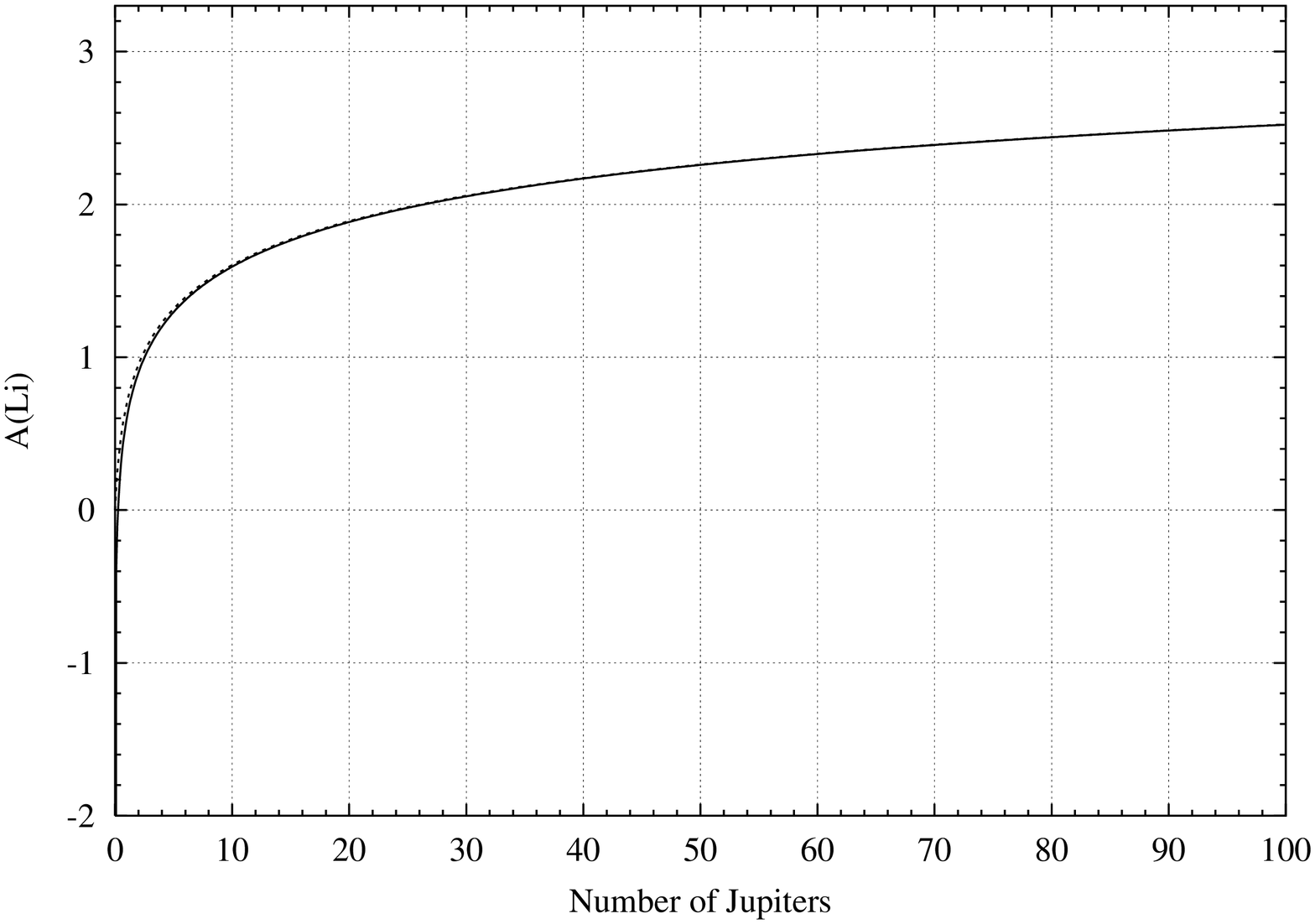}}&
\resizebox{0.47\hsize}{!}{\includegraphics[width=10cm,angle=-90]{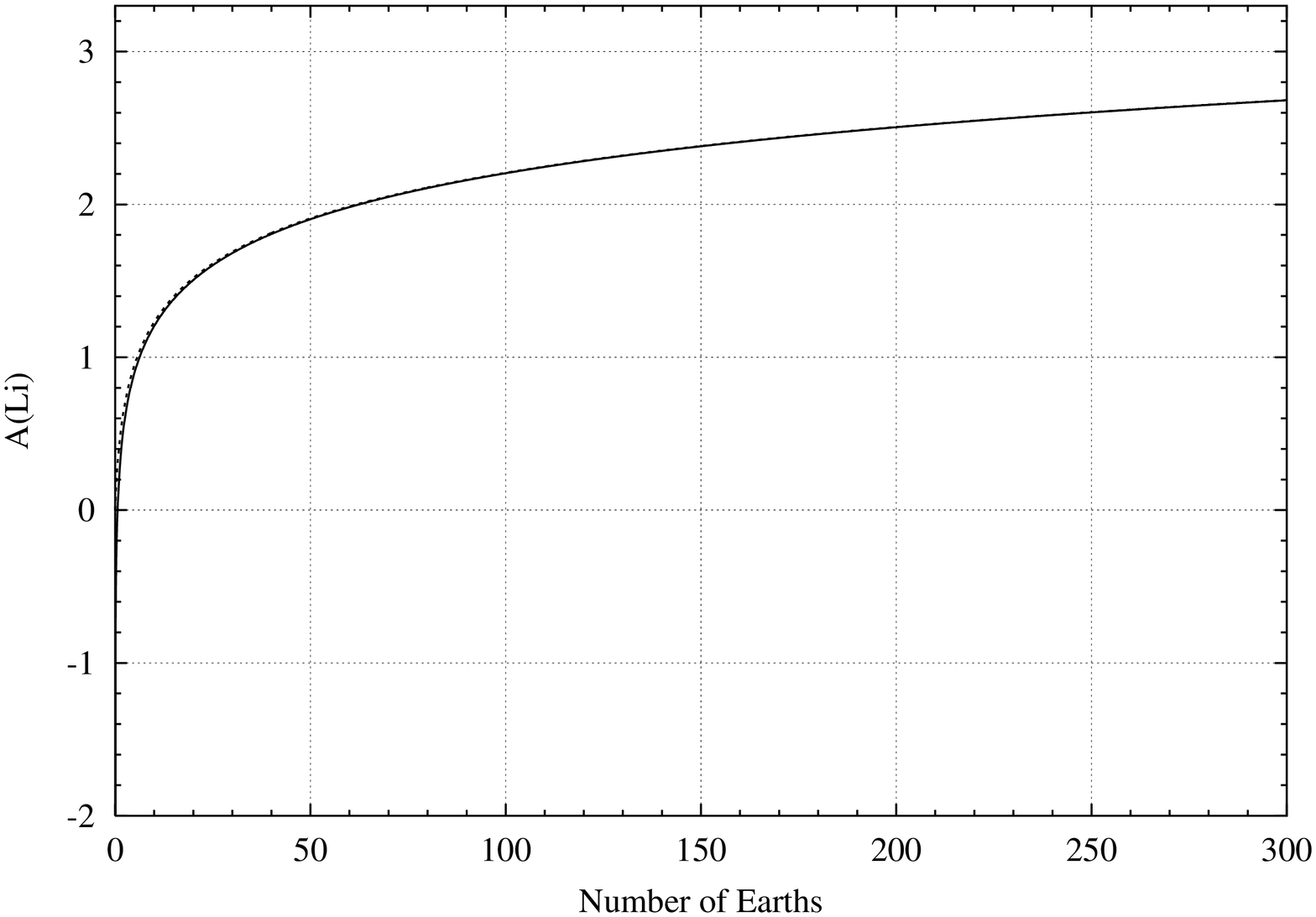}}\\
\resizebox{0.47\hsize}{!}{\includegraphics[width=10cm,angle=-90]{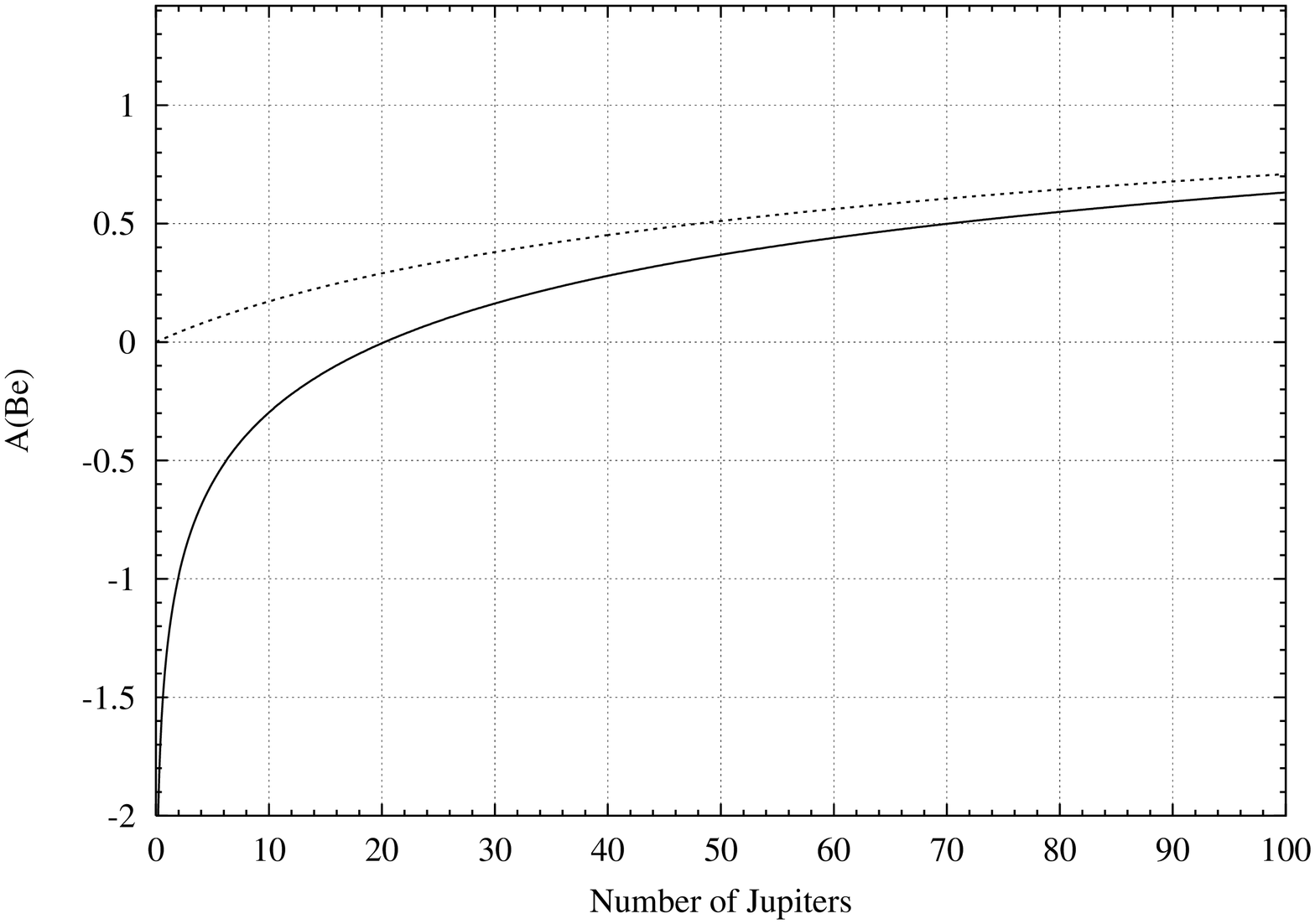}}&
\resizebox{0.47\hsize}{!}{\includegraphics[width=10cm,angle=-90]{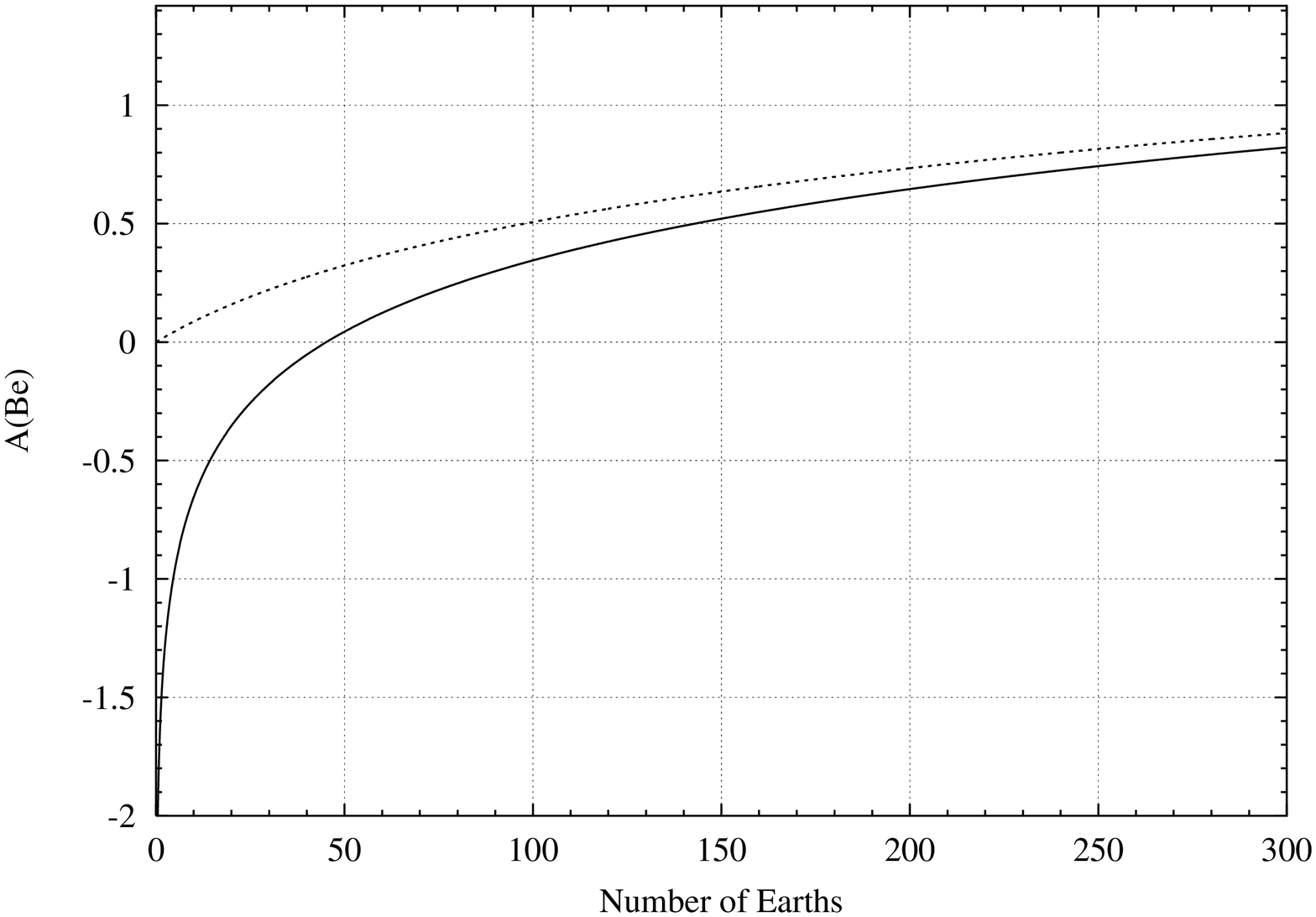}}\\
\end{tabular}
\caption{Li and Be abundances curves as a function of the mass and nature of the accreted material.
{\it Left} Top panel shows Li enrichment for two initial abundances $A(Li)=-10$ (solid line) and 
$A(Li)=0.0$ (dotted line) as a function of the number of Jupiters engulfed.
In the bottom panel similar curves are shown for the Be enrichment.
{\it Right.} Li and
Be enrichment curves for chondritic--like engulfed material given in Earth mass.
In all cases, a convective zone of 0.5$M_\odot$ has been assumed. See text for more details.}
\label{fig:pollution}
\end{figure*}

In summary, the complete absence of Be in the atmosphere of
the Li-rich giants suggests that it is very unlikely (from the physical
point of view) that the Li that we see is that preserved from the main-sequence phase.
Our results strongly favor the BCP mechanism proposed by Sackmann \& Boothroyd (\cite{sacboo99})
as the source of Li enrichment. However, as pointed out by Charbonnel \& Balachandran (\cite{chabal00}),
this scenario still presents some caveats.
A similar conclusion was drawn by
Castilho et al. (\cite{casetal99}), who 
derived $A(Be)$ for three giants, two Li-rich (HD787 and HD148650)
and another Li-normal (HD220321). They found that these stars had depleted more the 90\% of their 
Be (assuming an initial $A(Be)=1.42$) regardless of their Li-rich or Li-normal nature.
All of the three giants studied by Castilho et al. (\cite{casetal99}) have been re-observed 
in this work. Although the qualitative results are the same, the absolute values of abundances
are different, as while Castilho et al. still find some Be in the atmosphere of these 3
stars, we found no trace of such element. A possible explanation might come from the difference in
resolution and sensitivity of the spectra collected with
CASPEC and those taken to UVES, which has roughly 2 times more resolving
power and a $S/N$ twice as good.
Another more plausible explanation is a difference in the 
linelists. Castilho et al. added an {\it ad-hoc} Fe \textsc{i} at 3130.995\AA\ in order to fit
the Be region of two dwarfs, $\alpha$ Cen A and $\alpha$ Cen B used as reference stars.
In our case, we found it more appropriate to calibrate our oscillator strength using Arcturus as 
a reference star. As discussed in Sec. 3.2, this procedure lead to a good fit without any need
to include extra iron lines.

\subsection{Engulfment episode?}

The observational fact that most of the Li-rich giants are associated with IRAS sources led to de la Reza et
al. (\cite{delareza96}) and de la Reza et al. (\cite{delareza97}) to propose a scenario linking the Li excess
to the infrared properties.  They suggest that the Li enrichment produced by an internal Li production
mechanism based on the Cameron-Fowler  mechanism will also trigger the formation of a circumstellar shell. The mass loss
suddenly stops and the CS detaches making the star move in the IRAS color-color diagram  from zone I to II, II
to III and finally back to I again in which
they call {\it the Li-cycle}. (see e.g., Figure 1 of de la Reza et al.~\cite{delareza97}). 
In parallel to the expansion of the circumstellar shell, 
fresh Li is destroyed by the very same mechanism which has created it, namely, the deep mixing. 
Thus the actual location of the star in the IRAS color--color diagram depends both on the
difference between the time-scale of the expansion of the circumstellar shell and the 
time-scale for Li depletion.
In spite of being attractive, this scenario does not explain how the deep mixing and the mass loss
phenomena are connected. Also, precise predictions for the photospheric abundance pattern 
($^7$Li, $^9$Be, $^{12}$C/$^{13}$C, for instance) as a function
of time are not given.

Siess \& Livio (\cite{siesslivio99})
studied in detail the implications of the engulfment of a giant  planet or a brown dwarf by a giant stars.
They show that such an event will be accompanied by several
observational signatures such as production and posterior detachment of 
circumstellar shell, light element enrichment, spin-up and X--ray activity enhancement.
Concerning the circumstellar shell and its detachment, it
occurs as consequence of the thermal pulse caused by the accreted material (i.e., engulfed object).
Their calculations show that the higher the accretion rate, the larger is the increase in the 
mass-loss rate and the more evolved the star, the easier it is to remove material from its surface
due to a lower surface gravity. Thus the Li-cycle as proposed to de la Reza et al. 
(\cite{delareza96}, \cite{delareza97}) is naturally explained as the consequence of the engulfment phenomena.

%Explicar e passar as abundacias e no final dizer que o model tem difciucldades em epxlicar Li=3

%Quanto deveria enriquecer a estrela em Li e em Be? O excesso de Li e detectavel? e o de
%Be?
Another important observational signature of a engulfment episode 
is the light element enrichment mainly of $^7$Li and $^9$Be. 
Siess \& Livio (\cite{siesslivio99}) acknowledges already the fact in order to explain 
Li abundances as high as $A(Li)\sim2.5$ based on their engulfment scenario either i) 
the amount of mass accreted was unlikely high (i.e., several planets) or ii) the Li content of a brown dwarf
is extremely high (Li mass fraction would be typically 10 times higher in the brown dwarf than in the stellar
envelope). In order to overcome this problem,
they claim that accreted material will also trigger the cool bottom burning discussed above which will
help to increase the Li content.

%1-Calculo de quanto Be e esperado dado que o $^7$Li e de tanto (importante)
In light of the total absence of Be in these stars (c.f. Sec. 4.1),
can the hypothesis of a Li enrichment
due to accretion of an external Li-rich material be ruled out?
In Figure~\ref{fig:pollution} we show the final abundances of Li and Be expected for a giant star
with a convective zone of 0.5$M_\odot$ and for two initial Li and Be abundances
as a function of the accreted mass. The accreted material is either of Jupiter- (left) or
chondritic-like material. In the case of a Jupiter--like material, meteoritic abundances for
Li and Be have been assumed (Anders \& Grevesse \cite{andersgrevesse89}) 
whereas for the chodritic-like material Li- and Be-to-iron ratios have been assumed to be
meteoritic (Anders \& Grevesse \cite{andersgrevesse89}) and
the mass fraction of iron to be 19\%.

Looking at Figure~\ref{fig:pollution}, we see that 
in order to match the Li abundance of a 
very Li-rich giant ($A(Li)\sim2.5$), an ordinary post-diluted giant has to
accrete $\sim100$ Jupiters, i.e., $0.1M_\odot$, regardless the initial Li abundances 
($A(Li)=-10$ $A(Li)=0.0$).
As far as the Be is concerned, for
an accreted mass of $0.1M_\odot$, the final $A(Be)$ is expected to be 0.6-0.7 depending on the
initial Be content. For an accreted mass of 10 Jupiters, the Li and Be abundances are expected to be
1.5 and $-0.3$, respectively.  Now, if the composition of the accreted material is chondritic-like, then
we would need about 40 Earth-masses of chondritic-like material in 
order to enrich the envelope of giant star to the level
of $A(Li)\sim2.0$ which would bring the $A(Be)$ to about 0.0.
It is interesting to note however that if close-in giant planets are
super-metallic (e.g. due to hydrogen evaporation - 
Vidal-Madjar et al.~\cite{vidmaj03}), then a lower, and eventually 
reasonable number of accreted Jupiter-like planets would 
be needed for the star to achieve a high Li-abundance.

In Figure~\ref{fig:synth} we show the spectra obtained for four of the program stars overplotted with the
spectral synthesis for different Be abundances, namely, $A(Be)$= 0.0,
-1.0, -1.5, -2.0, -5.0 (no Be). We see that a $A(Be)$
abundance of $\sim0.0$ (corresponding to a $A(Li)=1.5$ and an accreted mass of 10 Jupiters)
would be easily detected in our observations according to Figure~\ref{fig:synth} 
($A(Be)\sim0$ corresponds to $A(Be)\sim-1.2$).  Thus, our 
newly derived $A(Be)$ show that it is
unlikely that the Li-rich giants were enriched via engulfment episodes.
Moreover, the absence of Be suggest that a Cameron-Fowler mechanism is behind the
Li enrichment in these stars. However the IR excess observed for a few Li-rich and Li-normal
stars still remained to be fully explained.

%Engulfment vai ocorrer de todos os modos

The typical radius of a solar mass red giant is of about 70$R_\odot$ or $\sim$0.3 AU which corresponds to roughly
the size of the orbit of Mercury, i.e., to an orbital period of about 60-100 days.
Given the orbital period distribution of the known extrasolar planets, 
we see that $\sim30$ per cent of the known extra-solar planets are going to surfer the inexorable 
fate of being engulfed by their own sun when 
it reaches the red giant branch.  Therefore {\it engulfment does occur}. 
Our results rather than rule out 
the engulfment phenomenon itself, it only says that, given the $^7$Li and $^9$Be abundances shown here,
engulfment as the main source of Li enrichment is ruled out.

%Porque algumas estrelas passam pelo Cemeron Fowler e outras nao?

Finally, through out the paper we have argued that it is very likely that the Cameron-Fowler mechanism is 
responsible for the Li enrichment observed in some giant stars. However what triggers the Cameron-Fowler 
mechanism is not clear. Also, why is it ignited in some stars and not in others? These are important open questions
which are beyond the scope of this paper.

\section{Conclusions}

In this paper we have analyzed the Be abundances for a sample of Li-rich giant stars with the
aim of testing the engulfment hypothesis suggested by Siess \& Livio (\cite{siesslivio99}).
The main results obtained can be outlined as follows:

\begin{itemize}

\item[-] {\bf No Be was found in any of the observed stars, 
regardless their nature Li-rich or Li-normal}.
The absence of Be is real and it is not an artifact of the spectral syntheses as
shown by the good fit achieved for Arcturus (Fig.~\ref{fig:arcturus})
and for the Li-rich and Li-normal giants (Fig.~\ref{fig:synth}).

\item[-] {\bf The Cameron-Fowler mechanism is favored.} The presence of Li and the absence of Be strongly suggests
that the Li enrichment is probably result of dredge-up of fresh Li produced in the interior of 
these stars
by the extra-mixing mechanism proposed by Sackmann \&  Boothroyd (\cite{sacboo99}). 
However, the fact that the model fails to predict $^{12}$C$/^{13}$C observed indicates that it still
requires improvement.

\item[-] {\bf Engulfment cannot explain the presence of Li and absence of Be}.
Using simple dilution arguments, we show that the accreted mass necessary to produce the Li-rich
enrichment observed in the Li-rich giants would also produce a Be enrichment detectable by our observations.
Thus our newly derived Be abundances rule out engulfment as the {\it sole} source of the Li observed
in the Li-rich giants.
 
%Finally, the differences found between our Be abundances and those derived in
%Castilho et al. (\cite{casetal99}) remain to be fully understood. However, 
%no matter the absolute values of $n(Be)$ for the stars listed in Table~\ref{tab:param} might be, 
%their spectra do not show any trace of Be. Even if the results of Castilho et al. turn to be
%the correct abundance values, the discussion and the conclusion drawn in this paper 
%remain intact.

\end{itemize}

\begin{acknowledgements} We would like to warmly thank the Paranal Observatory
 staff, specially the UT2/Kueyen service mode observers, for carrying out these
 observations in an efficient and competent way. Support from Funda\c{c}\~ao
 para a Ci\^encia e Tecnologia (Portugal)  to N.C.S. in the form of a
 scholarship is gratefully acknowledged. J.D.N.Jr. acknowledges the CNPq grant
 PROFIX 540461/01-6. J.R.DM has been supported by continuous grants from the
 CNPq Brazilian Agency. We are grateful to Dr. Guillermo Gonzalez who has
 kindly provided
 the Arcturus spectrum used in this work. Finally, we thank the anonymous
 referee whose comments helped to improved the paper.

  \end{acknowledgements}

\end{document}